\documentclass[sigconf]{acmart}
\AtBeginDocument{%
  }

\copyrightyear{2026}
\acmYear{2026}
\setcopyright{cc}
\setcctype{by}
\acmConference[ICSE-SEIS '26]{2026 IEEE/ACM 48th International Conference on Software Engineering}{April 12--18, 2026}{Rio de Janeiro, Brazil}
\acmBooktitle{2026 IEEE/ACM 48th International Conference on Software Engineering (ICSE-SEIS '26), April 12--18, 2026, Rio de Janeiro, Brazil}
\acmPrice{}
\acmDOI{10.1145/3786581.3786932}
\acmISBN{979-8-4007-2424-4/2026/04}

\usepackage[utf8]{inputenc}

\PassOptionsToPackage{hyphens}{url}\usepackage{hyperref}

\usepackage{csquotes}

\usepackage{graphicx}
\graphicspath{{images/}}

\usepackage{multirow}

\usepackage{makecell}

\usepackage{array}
\renewcommand{\arraystretch}{1.1}

\usepackage{enumitem}
\setlist[itemize]{leftmargin=20pt}

\usepackage{dblfloatfix}

\usepackage{footnote}
\makesavenoteenv{tabular}
\makesavenoteenv{table}

\usepackage{subcaption}

\usepackage{listings}
\usepackage{xcolor}

\begin{document}

\title[On the Effectiveness of Proposed Techniques to Reduce Energy Consumption in RAG Systems]{On the Effectiveness of Proposed Techniques to Reduce Energy Consumption in RAG Systems: A Controlled Experiment}

\author{Zhinuan (Otto) Guo}
\orcid{0009-0003-2963-6299}
\affiliation{
    \institution{Vrije Universiteit Amsterdam}
    \city{Amsterdam}
    \country{The Netherlands}
}
\email{guozhinuan2@gmail.com}

\author{Chushu Gao}
\orcid{0000-0003-1397-4536}
\affiliation{
    \institution{Software Improvement Group}
    \city{Amsterdam}
    \country{The Netherlands}
}
\email{chushu.gao@softwareimprovementgroup.com}

\author{Justus Bogner}
\orcid{0000-0001-5788-0991}
\affiliation{
    \institution{Vrije Universiteit Amsterdam}
    \city{Amsterdam}
    \country{The Netherlands}
}
\email{j.bogner@vu.nl}




\renewcommand{\shortauthors}{Guo et al.}

\begin{abstract}
The rising energy demands of machine learning (ML), e.g., implemented in popular variants like retrieval-augmented generation (RAG) systems, have raised significant concerns about their environmental sustainability.
While previous research has proposed green tactics for ML-enabled systems, their empirical evaluation within RAG systems remains largely unexplored.
This study presents a controlled experiment investigating five practical techniques aimed at reducing energy consumption in RAG systems.
Using a production-like RAG system developed at our collaboration partner, the Software Improvement Group, we evaluated the impact of these techniques on energy consumption, latency, and accuracy.

Through a total of 9 configurations spanning over 200 hours of trials using the CRAG dataset, we reveal that techniques such as increasing similarity retrieval thresholds, reducing embedding sizes, applying vector indexing, and using a BM25S reranker can significantly reduce energy usage, up to 60\% in some cases.
However, several techniques also led to unacceptable accuracy decreases, e.g., by up to 30\% for the indexing strategies.
Notably, finding an optimal retrieval threshold and reducing embedding size substantially reduced energy consumption and latency with no loss in accuracy, making these two techniques truly energy-efficient.
We present the first comprehensive, empirical study on energy-efficient design techniques for RAG systems, providing guidance for developers and researchers aiming to build sustainable RAG applications.
\end{abstract}

\ccsdesc[500]{Software and its engineering~Designing software}
\ccsdesc[500]{Software and its engineering~Software performance}
\ccsdesc[500]{Computing methodologies~Machine learning}

\keywords{green ML engineering, retrieval-augmented generation, energy consumption, latency, accuracy, controlled experiment}


\settopmatter{printfolios=true}

\maketitle

\section{Introduction}\label{s:intro}
Large language models (LLMs) as the most recent wave of machine learning (ML) have shown impressive performance in natural language processing tasks and beyond~\cite{chang_survey_2024}.
However, they also face several challenges, including hallucination~\cite{Ji_2023_Hallucination_nlp, xu2025hallucinationinevitableinnatelimitation}, outdated knowledge, and opaque or untraceable reasoning processes~\cite{gao2024retrievalaugmentedgenerationlargelanguage}.
Retrieval-augmented generation (RAG) has emerged as a popular technique to address shortcomings by integrating knowledge from external databases into the generation process~\cite{Ji_2023_Hallucination_nlp}.

As ML-enabled systems are increasingly adopted, more attention is also spent on their carbon footprint~\cite{Luccioni_2024, Bashir2024Climate}.
Due to factors such as increased technology adoption, cryptocurrency trends, and the rising demand for artificial intelligence, global data center electricity consumption is projected to range between 620 and 1,050 TWh by 2026~\cite{IEA2024Electricity}.
The growing popularity of RAG systems is adding to this consumption, as they are increasingly applied in diverse domains such as code generation~\cite{parvez2021retrievalaugmentedcodegeneration, zhou2022docprompting}, question answering~\cite{hu-etal-2022-logical, huang-etal-2021-unseen-entity}, and AI for science~\cite{wang2023retrievalbasedcontrollablemoleculegeneration}.
Their energy consumption and carbon footprint should therefore be a subject of concern.

While research on the environmental impact of ML-enabled systems has received more interest in recent years~\cite{Verdecchia2023GreenAI, Gezici2022}, the quality attribute (QA) of environmental sustainability has received little direct attention in RAG systems.
Existing research predominantly emphasizes other QAs, such as accuracy and latency, while largely neglecting energy efficiency.
For instance, \citet{wang2024searchingbestpracticesretrievalaugmented} investigated various RAG configurations but focused primarily on performance metrics like accuracy and latency.
Similarly, \citet{a_synthesis_green_tactics_ml_sys} have proposed 30 green architectural tactics for generic ML-enabled systems, but the practical applicability of these tactics within the context of RAG systems remains largely unexplored.
This underscores a clear research gap: studying how to reduce energy consumption as a core quality concern in RAG systems and systematically exploring trade-offs with other system attributes.
While some techniques have been proposed to improve the resource utilization or latency of RAG systems, no systematic study has examined the energy efficiency impact of RAG techniques to provide guidance to practitioners about their usage.

To close this gap, the primary objective of this research is to analyze the energy consumption of various techniques in RAG systems while also examining potential trade-offs with response latency and the accuracy of generated answers.
We conducted a controlled experiment~\cite{wohlin12} involving five techniques, some tested under different parameter settings, resulting in a total of nine configurations that we compared to a baseline industrial RAG system.
To the best of our knowledge, this is the first study to conduct an in-depth investigation into different energy techniques for RAG systems.
Our findings provide valuable insights for future research on the environmental sustainability of such systems and offer guidance on navigating trade-offs between energy usage, latency, and answer accuracy.

\section{Related Work}
In this section, we discuss related work in the area of ML energy efficiency and empirical studies about RAG systems.

\citet{a_synthesis_green_tactics_ml_sys} presented a synthesis of green architectural tactics for ML-enabled systems.
They compiled a catalog of 30 tactics, derived from an extensive review of Green AI literature and validated through an expert focus group.
These tactics span six categories, namely data-centric strategies, algorithm design, model optimization, model training, deployment, and management, aimed at reducing energy consumption and enhancing computational efficiency throughout the ML system lifecycle.
While their work provides valuable, actionable guidance for building sustainable ML-enabled systems, the proposed tactics are not specifically tailored to RAG systems, and evidence for their effectiveness in this context is missing.
In this study, we focus specifically on production-level RAG systems and empirically evaluate the impact of selected techniques on energy efficiency.

Using experiments to demonstrate that current LLM agents and RAG systems consume substantial amounts of energy, \citet{wu2025addressingsustainableaitrilemma} introduced the \enquote{Sustainable AI Trilemma}, highlighting the tensions between  AI capability, negative environmental impact, and digital inequality.
They reported that common optimization steps, e.g., query optimization or compression, drastically increase energy use with diminishing returns and that LLM-dependent methods consume orders of magnitude more energy than non-LLM alternatives.
While their paper crucially highlights these issues and inefficiencies within RAG systems, it stops at diagnosis and does not propose or study specific green techniques as a solution.
In contrast, our research focuses explicitly on evaluating proposed technique variations for designing green RAG systems.
 
Current techniques targeting RAG systems primarily focus on improving other key performance metrics, such as response time or reducing computational resource usage. 
In several cases, this can also lead to reductions in energy consumption.
For example, \citet{arefeen2024irag} examined the effect of threshold k selection in chunk filtering within the iRAG framework, demonstrating that a balanced selection of k enhances query efficiency and reduces computational waste.
While increasing k results in longer query processing and greater computational overhead due to processing more candidate chunks, effective chunk filtering mitigates this by minimizing unnecessary computation while preserving high recall.
However, there remains a lack of broad empirical evidence on the effectiveness and trade-offs of various energy-saving techniques specifically designed for RAG systems, especially for realistic systems from an industry context.

Similarly, \citet{Sakar_Emekci_2025} conducted a comprehensive evaluation of various RAG systems, identifying optimal configurations that balance critical performance metrics like response accuracy, token efficiency, runtime, and hardware utilization across diverse domains.
Their findings indicate that Reciprocal RAG, which generates and ranks multiple query variations to resolve ambiguity, achieved the highest similarity score but at the cost of significantly higher token usage and longer run times.
In contrast, the Stuff method, which simply stuffs all retrieved documents into a single prompt, was the fastest and most token-efficient approach but sacrificed response accuracy by not addressing query ambiguity.
However, their study does not offer any insights on the energy consumption of the methods, omitting a crucial dimension for a comprehensive system evaluation.
Our research puts energy consumption at the center and also studies potential trade-offs with latency and accuracy.

In a study from the medical domain, \citet{11101522} systematically evaluated the end-to-end performance of RAG systems on the RAGEval DragonBall dataset, employing retrieval metrics (recall, precision, MAP) and generation metrics (RAGAs, BERTScore) to compare cost-effective sparse and dense retrieval methods against commercial embeddings.
Their key finding was that the sparse method BM25 outperformed all dense and commercial embedding models in the retrieval stage and achieved the highest scores in most generation metrics.
While their research provides valuable insight into the effectiveness of open sparse retrieval methods like BM25, it is limited to a single technique and medical domain benchmark dataset.
In contrast, our research extends this investigation to a real-world production application by evaluating a RAG system adopted by our industry partner.
This allows us to explore the performance and challenges of these retrieval methods under practical operational conditions and constraints.

Overall, there is a lack of empirical evidence about techniques that directly optimize the energy consumption in RAG systems.
Despite the increasing popularity of RAG as a framework within ML-enabled applications, its energy consumption characteristics and techniques to improve them remain underexplored.
Most existing studies on energy efficiency focus on general ML-enabled systems rather than RAG.
While these more general techniques can still be of situational value for designing greener RAG architectures, e.g., by selecting energy-efficient algorithms~\cite{Kaack2022}, more specific green RAG techniques and evidence about their effectiveness and trade-offs are needed.
Our study aims to start closing this gap.

\section{Study Design}
To ensure both academic rigor and industrial relevance, we formed an academia-industry collaboration between our university and the Software Improvement Group (SIG), a software consultancy firm specialized in software quality and digital sustainability with roughly 160 employees.\footnote{\url{https://www.softwareimprovementgroup.com}}
We started by choosing and analyzing one of their suitable RAG systems.
After that, we together reviewed and discussed the scientific literature on RAG systems to identify and select suitable techniques for their system.
Finally, we conducted an industry-informed controlled experiment with selected techniques.
The primary objective of our research was to evaluate the effectiveness of current techniques in reducing energy consumption in RAG systems and to study potential trade-offs with relevant QAs.
Towards this end, our study addresses the following two research questions:

\begin{enumerate}
    \item[\textbf{{RQ}1}] How effective are proposed techniques for reducing the energy consumption of RAG systems?
    \item[\textbf{{RQ}2}] How does the application of these techniques impact accuracy and latency?
\end{enumerate}

To maintain a manageable scope for the trade-off analysis (RQ2), we focused on two QAs that are essential for the effectiveness and acceptable user experience of industrial RAG systems for question answering: the accuracy and latency of responses.
Therefore, understanding the trade-off between energy consumption and these two QAs is crucial. Unnecessary retrieval may increase latency and computational costs, while insufficient retrieval may lead to incomplete or incorrect answers~\cite{su2025fastbetterbalancingaccuracy}.
Thus, investigating the trade-offs that techniques cause among these quality concerns holds significant value for practitioners and researchers alike.

\subsection{Experiment Objects}
Research on techniques for improving the environmental sustainability of RAG systems is unfortunately still scarce, and selecting promising candidates as experimental objects based on academic literature is nontrivial.
Therefore, we also included techniques that have not been specifically conceptualized in the context of energy efficiency but seemed promising in terms of reducing energy consumption.
To guide our selection of techniques, we applied the following inclusion criteria:

\begin{itemize}
    \item \textbf{Potentially energy-saving:} The adopted techniques must either have been directly recommended to lower energy consumption or seem reasonable to assume energy-related benefits due to, e.g., reduced resource consumption.
    \item \textbf{Not tied to proprietary solutions:} If a technique is closely related to proprietary solutions from companies like OpenAI or Bedrock, it is difficult to generalize the experiment results, and it may not be locally implementable.
    \item \textbf{Suitability for the current environment:} Techniques must be compatible with the existing system architecture and experiment environment. For instance, our GPU provides approximately 24 GB of VRAM, which limits how many models we can run simultaneously. Techniques requiring several LLMs, such as LLM-based context compression, are therefore infeasible.
    \item \textbf{Ease of implementation:} The technique should be implementable in the experimental system with reasonable effort.
\end{itemize}

Using these criteria, we finally selected five techniques, which are shown below.

\textbf{T1 – Increase similarity threshold of pgvector queries:}
The threshold in pgvector\footnote{\url{https://github.com/pgvector/pgvector}} refers to the similarity score used to filter results when querying vector embeddings in PostgreSQL.
This score determines which documents are returned based on their similarity to the user query.
A higher threshold returns fewer, more relevant documents, thereby reducing the amount of context retrieved in a RAG system.
This reduction may lead to reduced energy consumption, as less data needs to be processed.
\citet{bulgakov2024optimizationretrievalaugmentedgenerationcontext} showed that filtering semantically incoherent documents via thresholding significantly improves retrieval quality while minimizing memory and compute overhead.
The original baseline threshold in the RAG system under study was set to 0.58.
To better understand the impact of this value, we evaluated 100 test queries and found that the mean similarity score was 0.78.
This provided a statistically grounded reference point for typical retrieval behavior and helped avoid arbitrary threshold selection.
Since real production data is typically significantly larger and more diverse than the test sample, using the mean similarity score as a reference is more reasonable for generalization, as it captures the central tendency of similarity scores over a broader range of queries.
Based on this, we chose additional thresholds at regular intervals around these two values to examine how varying similarity cutoffs affect energy consumption and the other QAs (0.58 as the baseline, 0.68, 0.78, and 0.88).

\textbf{T2 - Introduce lightweight reranking algorithm:}
Reranking typically serves as an enhancement technique to improve the accuracy of RAG systems~\cite{sharifymoghaddam2025rankllmpythonpackagereranking}.
Since GPU inference typically consumes more energy than CPU operations, reducing the load on the GPU can help to lower overall energy usage.
In our experiment, we applied a lightweight reranking method using BM25S~\cite{bm25s} to filter and prioritize candidate documents before they are processed by the frozen LLM.
This approach reduces the number of documents passed to the LLM, potentially decreasing energy consumption while maintaining performance.

\textbf{T3 - Reduce embedding sizes of the embedding model:}
Reducing the size of word embeddings can improve their efficient use in memory-constrained devices, benefiting real-world applications~\cite{raunak-etal-2019-effective}.
In our experiment, we replaced the default embedding model in the baseline system (e5-large-v2\footnote{\url{https://huggingface.co/intfloat/e5-large-v2}}, 1024 dimensions) with smaller variants from the same provider: e5-base-v2\footnote{\url{https://huggingface.co/intfloat/e5-base-v2}} (768 dimensions) and e5-small-v2\footnote{\url{https://huggingface.co/intfloat/e5-small-v2}} (384 dimensions).

\textbf{T4 - Apply an efficient vector search indexing strategy:}
The indexing strategy of the vector search in RAG systems impacts how quickly the search is performed but also which and how many documents are selected.
Previous studies indicate that indexing via Hierarchical Navigable Small World (HNSW)~\cite{HNSW_indexing} or Inverted File with Flat Quantization (IVFFlat)~\cite{IVFFlat_indexing} can enhance efficiency, particularly with binary embeddings~\cite{herranen2024sustainable}.
HNSW indexing constructs a graph-based structure that facilitates efficient and robust approximate nearest neighbor search by traversing hierarchical layers of nodes.
IVFFlat indexing divides the data into clusters and employs inverted lists to rapidly narrow down the search space.
Since both indexing methods are well-suited for fast approximate nearest neighbor searches in high-dimensional spaces and also supported by pgvector, we included them both in our experiment.

\textbf{T5 - Cache intermediate retrieval states via knowledge trees:}
If several queries show decent similarity with each other, caching the intermediate knowledge in memory for reuse can enhance system efficiency by reducing redundant computation.
To allow this, \citet{jin2024ragcacheefficientknowledgecaching} have proposed the RAGCache approach that uses special structures called \textit{knowledge trees}.
This allows new queries to skip recomputation for shared prefixes and has been shown to improve overall latency and throughput in RAG systems~\cite{lu2024turboragacceleratingretrievalaugmentedgeneration, liu2024cachegen, cheng2024large, yao2024cacheblend}.
We adopt a configuration in vLLM for this called \texttt{enable\_prefix\_caching}~\cite{kwon2023efficient}, which enables the key-value (KV) cache of existing queries.

\begin{figure*}[bh]
    \includegraphics[width=0.85\linewidth]{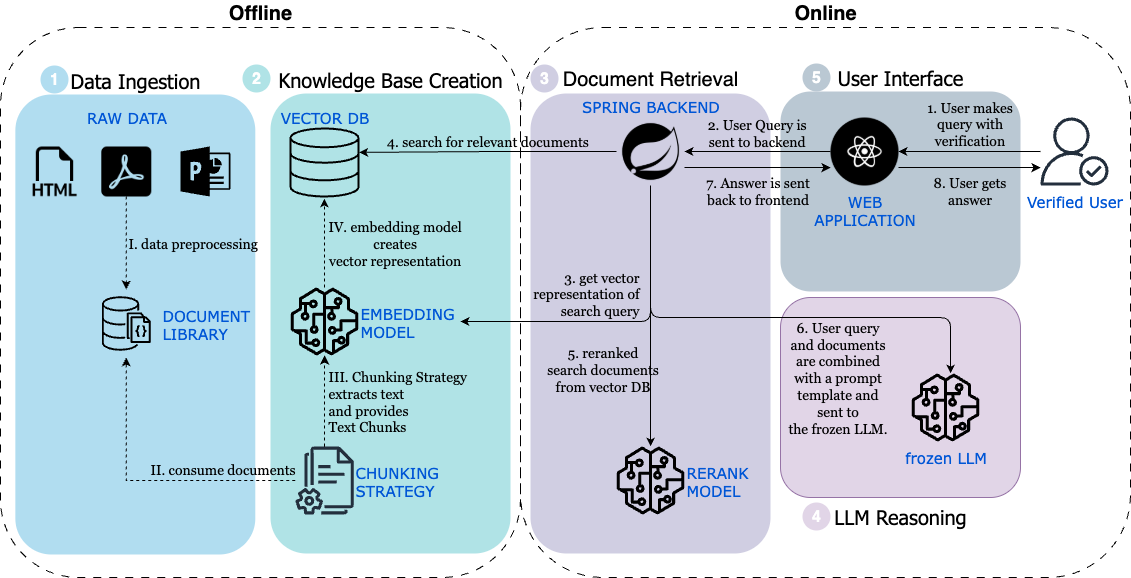}
    \caption{ChatRAG Software Architecture and Workflow}
    \label{fig:arg_architecture}
\end{figure*}

\subsection{Experiment Materials}
To ground the experiment design in real-world application requirements, we selected ChatRAG as our experiment system, a RAG system developed by SIG.
ChatRAG is a query-based chatbot that enhances its responses with insights from retrieved internal documents by feeding them directly into the initial stage of the generator~\cite{zhao2024retrievalaugmentedgenerationaigeneratedcontent}.
The specific architecture of ChatRAG is illustrated in Fig.~\ref{fig:arg_architecture}, while the detailed configuration of the baseline system and its components is provided in Table~\ref{tab:base_model}.

\begin{table}
    \centering
    \small
    \caption{ChatRAG Baseline System Characteristics}\label{tab:base_model}
    \begin{tabular}{p{1.4cm}p{2.3cm}p{3.7cm}}
        \textbf{Component} & \textbf{Technology} & \textbf{Explanation} \\
        \hline
        \hline
        Backend & Spring & Original ChatRAG code\\
        Embedding Model & E5 larger (V2) embedding model with 1024 dimension\footnote{\url{https://huggingface.co/intfloat/e5-large-v2}} & Open-source, supports different embedding sizes\\
        Frozen LLM & Llama 3.1 8B Instruct\footnote{\url{https://huggingface.co/meta-llama/Meta-Llama-3-8B-Instruct}} & Popular open-source LLM, supports quantization\\
        Vector Database & PostgreSQL 17\footnote{\url{https://www.postgresql.org/docs/current/release-17.html}} & Stable open-source database; same database as ChatRAG\\
        Reranking Model & -- & Discarded: not usable with experiment hardware\\
        \hline
        \hline
    \end{tabular}
\end{table}

To collect energy consumption data, we used Kepler\footnote{\url{https://sustainable-computing.io/}}, which attributes power usage to containers and Kubernetes Pods.
Kepler gathers real-time power consumption metrics from node components using Intel’s Running Average Power Limit (RAPL) for CPU and DRAM power, and the NVIDIA Management Library (NVML) for GPU power.
The collected and estimated container-level data is then stored using Prometheus.
To host the model in a Kubernetes environment with GPU resources, we adopted vLLM~\cite{kwon2023efficientmemorymanagementlarge}.
vLLM is a high-throughput and memory-efficient inference and serving engine for LLMs, supporting scalable deployment on Kubernetes.
With the assistance of the integration tool Helm\footnote{\url{https://helm.sh}}, we could efficiently deploy our models on Kubernetes.

To send and evaluate API calls to ChatRAG, we required an evaluation tool capable of measuring system performance.
We adopted LLMPerf\footnote{\url{https://github.com/ray-project/llmperf.git}}, which supports customization for issuing calls and collecting results related to latency and accuracy.
In our experiment, we extended this tool, which was originally based on the Ray distributed framework for emerging AI applications~\cite{moritz2018raydistributedframeworkemerging}, to support request handling, energy data collection, and accuracy tracking.
This extended version of the tool is publicly available on GitHub.\footnote{\url{https://github.com/KafkaOtto/rageval}}

\begin{figure}
    \includegraphics[width=\linewidth]{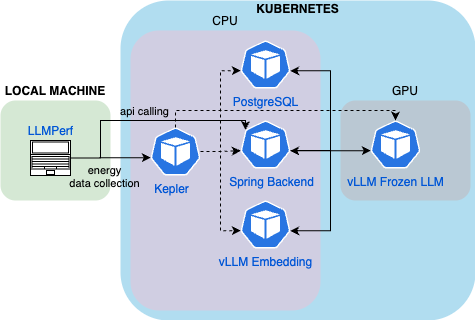}
    \caption{Experiment Environment and Workflow}
    \label{fig:experiment_setup}
\end{figure}

In real-world applications, direct energy measurements are often not feasible, especially when systems are hosted on virtualized public cloud infrastructure or if LLMs are accessed through enterprise APIs.
To enable accurate energy monitoring in a controlled environment, we therefore deployed the ChatRAG system variants on a server cluster in the Experiment Lab of our university,
a facility specifically designed for research on energy efficiency.
The machine used for the experiments was equipped with an NVIDIA GeForce RTX 4090 GPU (24 GB VRAM), an Intel Core i9-14900KF CPU (24 cores), and 128 GB of RAM.
The overall experiment setup is illustrated in Fig.~\ref{fig:experiment_setup}.
The frozen LLM was hosted on the GPU, while other components ran on CPUs.
A local machine with LLMPerf sent API calls to the backend pod and collected energy data from the Kepler pod.

\subsection{Experiment Variables}
The primary \textbf{independent variable} in our experiment is the system variations created by applying each selected technique.
Some techniques also included multiple configurations.
Specifically, T1 had three configurations with the similarity thresholds 0.68, 0.78, and 0.88 (baseline: 0.58).
T3 involved two configurations based on embedding sizes: 768 and 384 (baseline: 1024).
T4 included two configurations based on indexing strategies in pgvector: HNSW and IVFFlat (baseline: no indexing strategies).
T2 and T5 each consisted of a single configuration.
In total, that led to 9 technique configurations plus the baseline, i.e., 10 treatments in total.
The \textbf{dependent variables} for RQ1 were energy consumption (J), while they were accuracy (\%) and latency (s) for RQ2.

\subsection{Dataset}
Several criteria needed to be considered for selecting a benchmark dataset.
Firstly, the dataset had to be manageable, as very large-scale datasets were not suitable for our cluster.
For example, the KILT benchmark~\cite{petroni2021kiltbenchmarkknowledgeintensive}, designed for intensive language tasks, contains raw knowledge sources of approximately 35 GB, which would not be suitable for our experiment.
Secondly, the dataset had to be rich and cover a diverse set of topics, as dataset diversity can make the results more convincing.
Based on these criteria, we selected the CRAG dataset~\cite{crag_benchmark}, which contains a manageable total of 2,706 questions.
It also includes a diverse range of question types, such as comparison and multi-hop questions, which are reasonably close to real-world challenges commonly encountered in practical applications.
CRAG highlights the inherent trade-offs between accuracy and latency across various state-of-the-art RAG systems, offering useful context for evaluating ChatRAG’s performance.
The CRAG benchmark employs a structured scoring method to assess the quality of responses generated by RAG systems, categorizing each answer as perfect, missing, or incorrect.
Following this approach, we implemented a similar labeling process.
In our evaluation, we relied on an LLM-as-Judge approach~\cite{gu2025surveyllmasajudge} to validate the correctness of the final answers, excluding responses such as \enquote{I don’t know}.
LLM-as-Judge approaches provide efficient ways to evaluate query responses that go beyond simple requests like multiple-choice questions and have proven remarkably effective for many benchmarks, matching even crowdsourced human evaluations~\cite{Zheng2023}.
We employed DeepSeek-V2~\cite{deepseekai2024deepseekv2strongeconomicalefficient} as our LLM judge due to its low cost and competent question-answering capabilities.
The details of this evaluation and the used prompts are available in our replication package\footnote{\url{https://anonymous.4open.science/r/green-rag-techniques-experiment/running/README.md}}.


\subsection{Experiment Execution}
Before executing the experiment, we first had to prepare the CRAG documents for pgvector.
For the chunking strategy, we used the \texttt{TokenTextSplitter} of Spring AI.\footnote{\url{https://docs.spring.io/spring-ai/docs/current/api/org/springframework/ai/transformer/splitter/TokenTextSplitter.html}}
Due to the maximum token limit of 512 in the downstream embedding model, we set the chunk size window to 320, the same value as used in ChatRAG.
This resulted in a total of 501,916 split chunks.
Among these, 10 chunks were rejected by the embedding model because their token count exceeded the 512-token limit.
We chose to disregard these errors, as the overall number of chunks remains manageable.

For a test run of 1,335 queries, the total elapsed time was approximately 2 hours and 26 minutes.
We observed that it took around 5 minutes for the CPU to reach a stable usage level.
Since fluctuating CPU utilization could act as a confounder for measuring energy consumption, we therefore introduced a warm-up period to ensure that each trial was conducted under stable conditions.
The dataset was divided into two parts: a warm-up dataset consisting of 100 random queries and an experiment dataset containing the remaining queries.
The warm-up dataset was executed before each trial to stabilize the environment.
Additionally, we introduced a cool-down period of 5 minutes after each trial to allow system resources to reset.
The used dataset is available on Zenodo.\footnote{\url{https://doi.org/10.5281/zenodo.16569517}}

Furthermore, each experiment configuration, i.e., 9 technique variations + 1 baseline, was tested over a period of 10 trials to ensure the reliability of measurements and to limit the influence of potential random confounders.
A second reason for doing this was that we used the default configuration of Llama 3.1 8B Instruct, which employs a non-deterministic decoding strategy with a non-zero temperature and sampling enabled.
As a result, the model may produce slightly different outputs across runs for the same input.
While this increased variability was beneficial for external validity, the 10 trials per treatment were required to average out potential suboptimal generations.
We therefore report the average accuracy of all treatment runs.
In total, we ran 100 experiment trials.
Given that a single experiment trial took approximately 2 hours and 26 minutes, the total time required to complete the experiment was around 240 hours, i.e., slightly over 10 days.
This substantial runtime also explains why we could only consider a single dataset.

\subsection{Data Analysis}
After we collected all the data from the experiments, we assessed the normality of each data subset using the Shapiro-Wilk test~\cite{Hanusz_Tarasinska_Zielinski_2016}.
Since all datasets in our experiment passed the normality test, we used the t-test~\cite{student1908probable} to evaluate the significance levels.
Based on our hypotheses, we used a one-tailed test for energy consumption and a two-tailed test for performance.
Since multiple hypotheses were tested simultaneously, we had to guard against the multiple comparisons problem, where the probability of incorrectly rejecting at least one true null hypothesis (Type I error) increases with the number of tests performed~\cite{BARNETT20222331}.
To mitigate this, we applied the Holm-Bonferroni correction~\cite{holm1979simple}, which adjusts the p-values to control the family-wise error rate.
After correction, an adjusted p-value < 0.05 indicates a statistically significant difference, while a value > 0.05 suggests no significant difference.
For statistically significant differences, we calculated Cohen’s d~\cite{cohen2013statistical} to estimate the effect size.
According to Cohen’s guidelines, values between 0.0 and 0.2 indicate a negligible effect, 0.2 to 0.5 a small effect, 0.5 to 0.8 a moderate effect, and 0.8 and above a large effect.
Additionally, we calculated mean values for each configuration with a significant effect and then computed the percentage change relative to the baseline (control).

\section{Results}
\label{s:results}
In this section, we present our quantitative experiment results according to the research questions.
While we provide some interpretation for unexpected results, more explanations can be found in the discussion section.

\subsection{Reducing Energy Consumption (RQ1)}
\begin{figure}
    \centering
    \begin{subfigure}{\linewidth}
        \includegraphics[width=\linewidth]{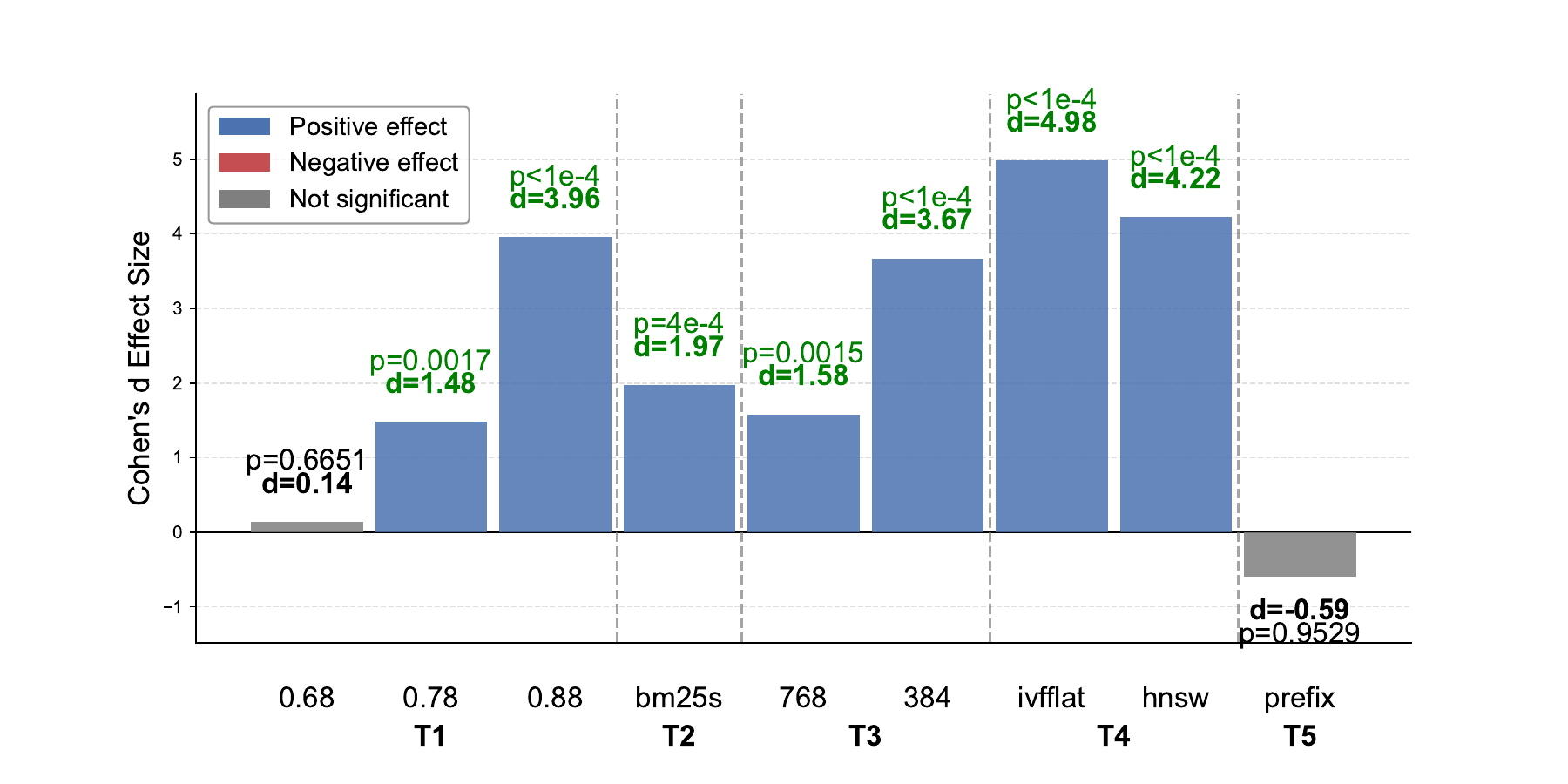}
        \caption{T-Test Results for Energy Consumption}
        \label{fig:t_test_energy}
    \end{subfigure}

    \vspace{0.0em}

    \begin{subfigure}{\linewidth}
        \includegraphics[width=\linewidth]{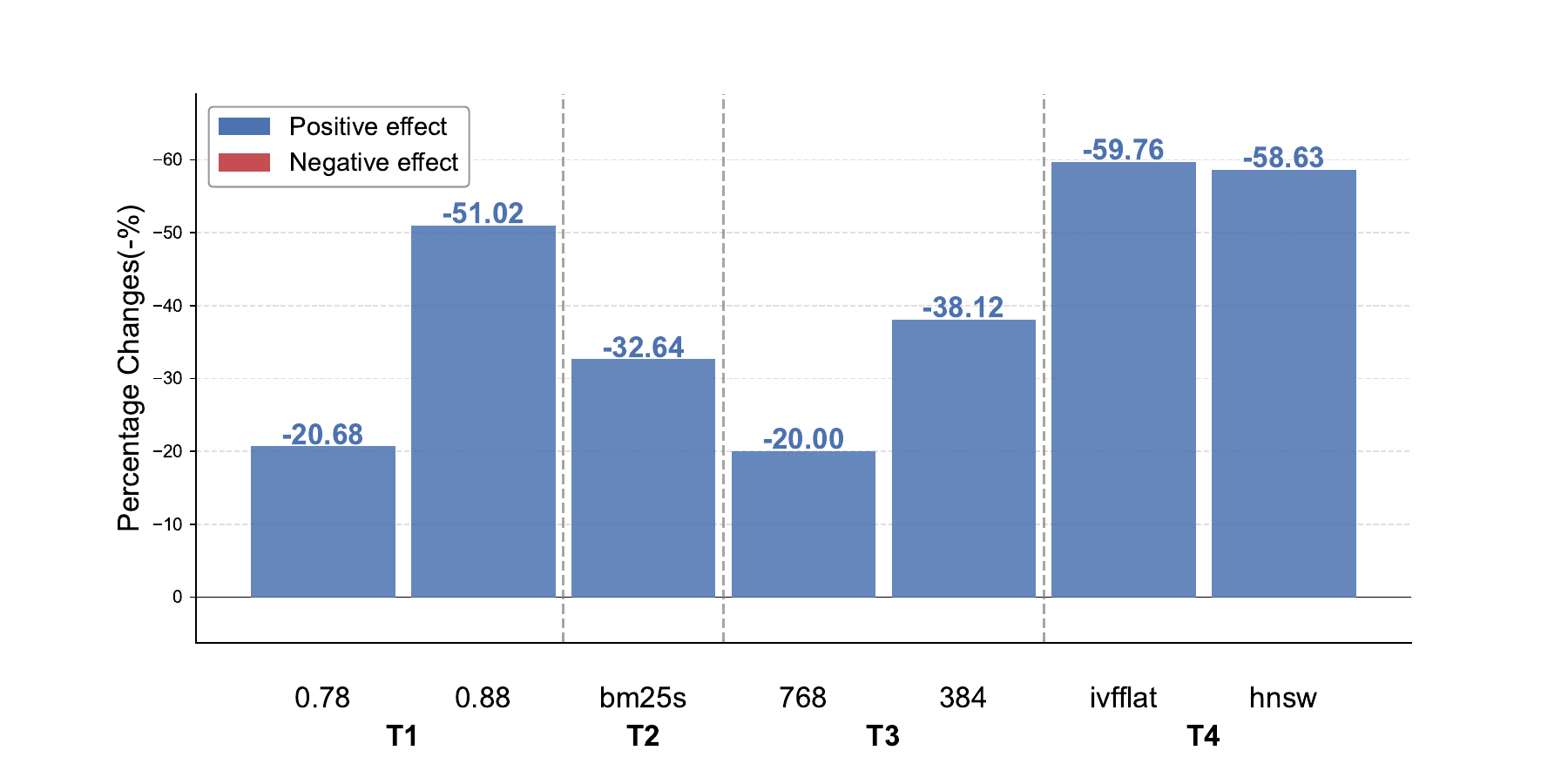}
        \caption{Relative Energy Difference to Baseline for Significant Techniques}
        \label{fig:percent_change_energy}
    \end{subfigure}

    \caption{Energy Consumption Experiment Results}
    \label{fig:energy_results}
\end{figure}

Based on the t-test results in Fig.~\ref{fig:t_test_energy}, we found that 7 of the 9 technique configurations had a statistically significant impact on the system’s energy consumption.
The exceptions were T1 with a similarity threshold of 0.68, for which the small reduction was not significant, and T5 (enabling caching prefixes), which even led to slightly increased energy consumption.
For the significant techniques, all effects were very large (smallest Cohen's $d$ of 1.48), with some even reaching a Cohen's $d$ of 4.0 and above.

For the different thresholds of T1, we found that the difference between 0.68 and the 0.58 baseline was minor.
However, as the threshold increased to 0.78 and 0.88, energy consumption decreased significantly, with the 0.88 threshold showing the greatest reduction.
This outcome aligns with our expectations, as the average similarity score for the test data was approximately 0.78.
Therefore, thresholds of 0.58 and 0.68 retrieved documents with fairly close similarity scores, whereas thresholds of 0.78 and above filtered out more documents, reducing the computational load.
As shown in Fig.~\ref{fig:percent_change_energy}, using a threshold of 0.78 resulted in a 20.7\% reduction in energy consumption, while increasing the threshold to 0.88 achieved an impressive 51.0\% reduction.

\begin{figure}
    \includegraphics[width=\linewidth]{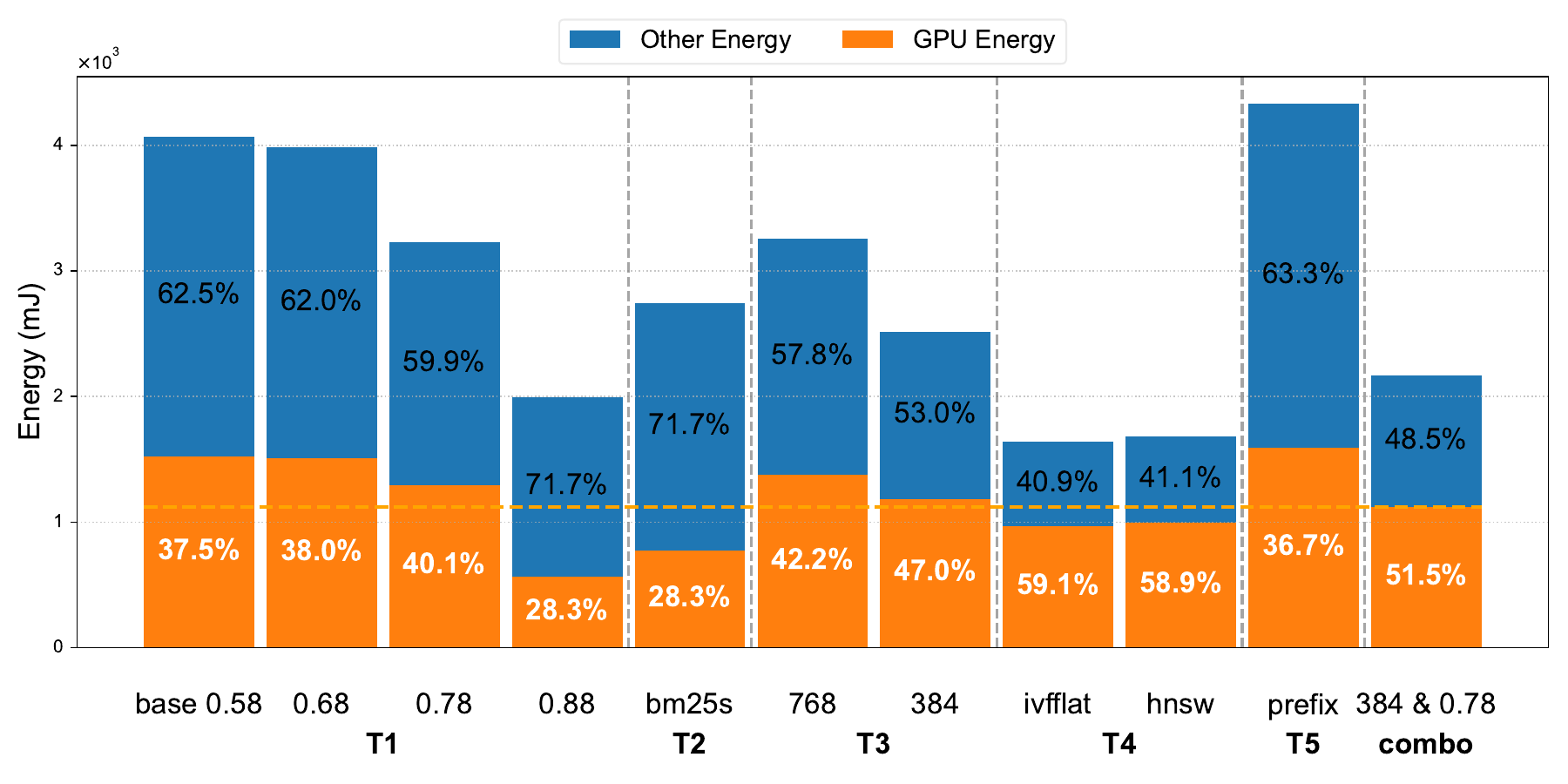}
    \caption{GPU Contribution to Total Energy Consumption.}
    \label{fig:energy_gpu_percentage}
\end{figure}

For the lightweight reranker T2, we observed a 32.6\% reduction in energy consumption.
Interestingly, as shown in Fig.~\ref{fig:energy_gpu_percentage}, both applying T1 with a threshold of 0.88 and T2 led to the same GPU energy usage percentage, with 28.3\% of the total energy, the lowest of all experiment configurations.
While the numerical similarity may be coincidental, the results suggest that both a high retrieval threshold and an effective reranking method can help filter irrelevant inputs early on.
This reduces the workload of the LLM, which is typically the most energy-intensive component of the system.

Regarding T3 (reducing the embedding size), both dimensions of 768 and 384 resulted in notable decreases in energy consumption (20.0\% and 38.1\%).
Since the embedding size defines the dimensionality of vector representations, smaller dimensions generated smaller vectors, which required less memory and computational resources and was therefore beneficial for energy use in practice.

Among all evaluated techniques, T4 yielded the most substantial energy reductions, with both indexing methods achieving approximately a 60\% decrease.
Indexing likely enabled faster and more efficient similarity searches, which not only lowered query latency and CPU usage for the database itself but also reduced overall system resource contention.
Additionally, the indexing strategies also reduced the number of fetched documents that were passed on to the LLM, which likely contributed to achieving the largest energy reductions among all evaluated techniques.

For T5, we enabled prefix caching on the frozen LLM during the generation stage.
However, no significant differences in energy consumption or performance were observed, and energy use even slightly increased compared to the baseline.
As shown in the Llama pod logs (Listing~\ref{lst:llama_logs}), cache utilization remained very low, with GPU hit rates of only $\sim$1.1\% and no hits on the CPU.

\begin{lstlisting}[caption={Llama Pod Logs for T5}, label=lst:llama_logs]
INFO 06-28 02:52:53 metrics.py:471] Prefix cache hit rate: GPU: 1.09%, CPU: 0.00%
INFO 06-28 02:52:58 metrics.py:455] Avg prompt throughput: 0.0 tokens/s, Avg generation throughput: 54.8 tokens/s, Running: 1 reqs, Swapped: 0 reqs, Pending: 0 reqs, GPU KV cache usage: 23.7%, CPU KV cache usage: 0.0%.
\end{lstlisting}

This limited effectiveness is likely due to the short prompt lengths and minimal overlap between retrieved documents across queries.
To quantify query similarity, we applied TF-IDF vectorization~\cite{Rajaraman_Ullman_2011} combined with cosine similarity.
The average similarity score was 0.0294 (standard deviation 0.0437), indicating that most queries differ substantially.
As a result, the prefix cache was rarely reused, leading to negligible improvements.
In contrast, systems with higher query similarity might benefit more from prefix caching.

\subsection{Trade-off With Latency (RQ2a)}
\begin{figure}
    \centering
    \begin{subfigure}{\linewidth}
        \includegraphics[width=\linewidth]{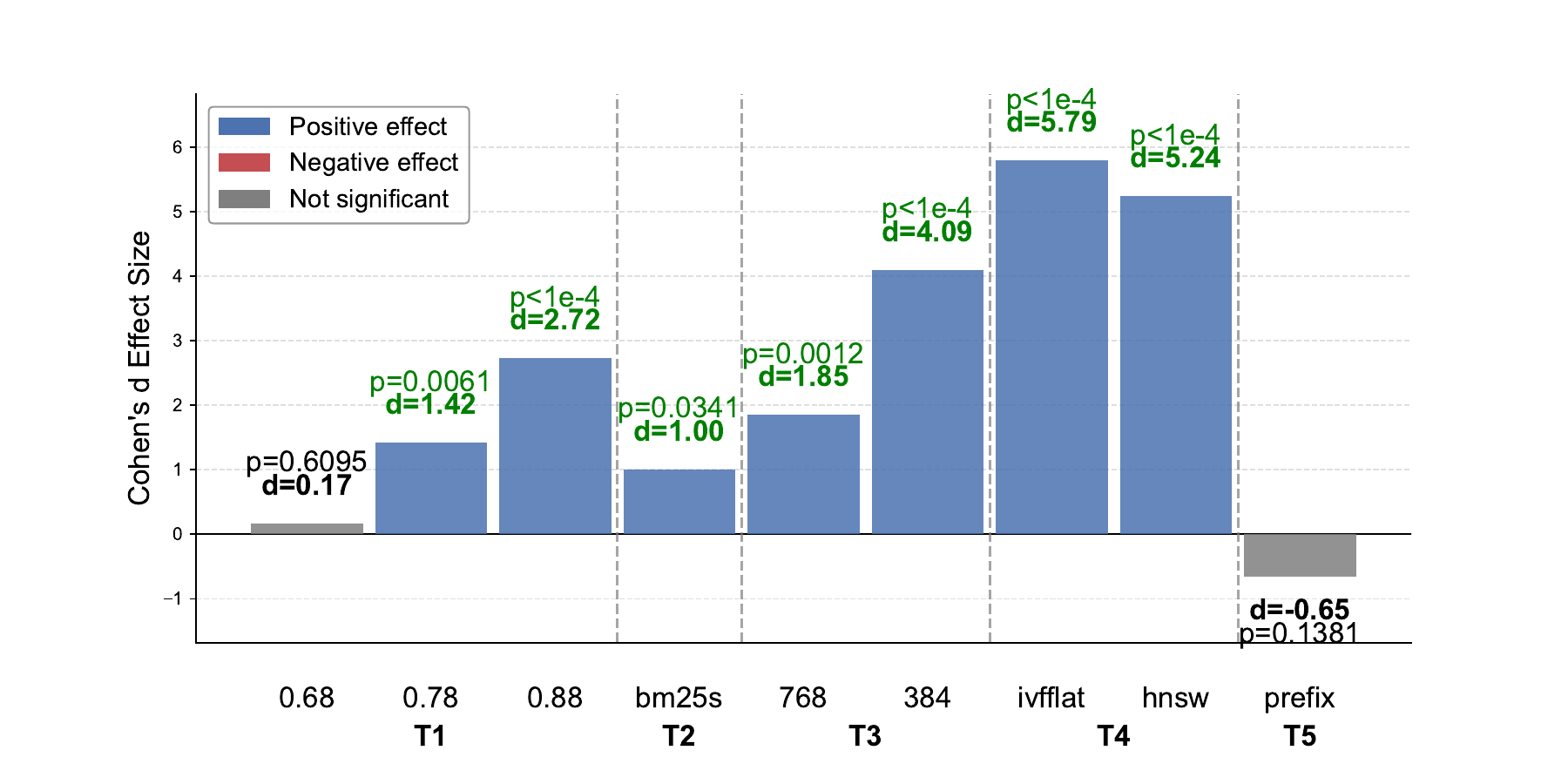}
        \caption{T-Test Results for Latency}
        \label{fig:t_test_latency}
    \end{subfigure}

    \vspace{0.0em}

    \begin{subfigure}{\linewidth}
        \includegraphics[width=\linewidth]{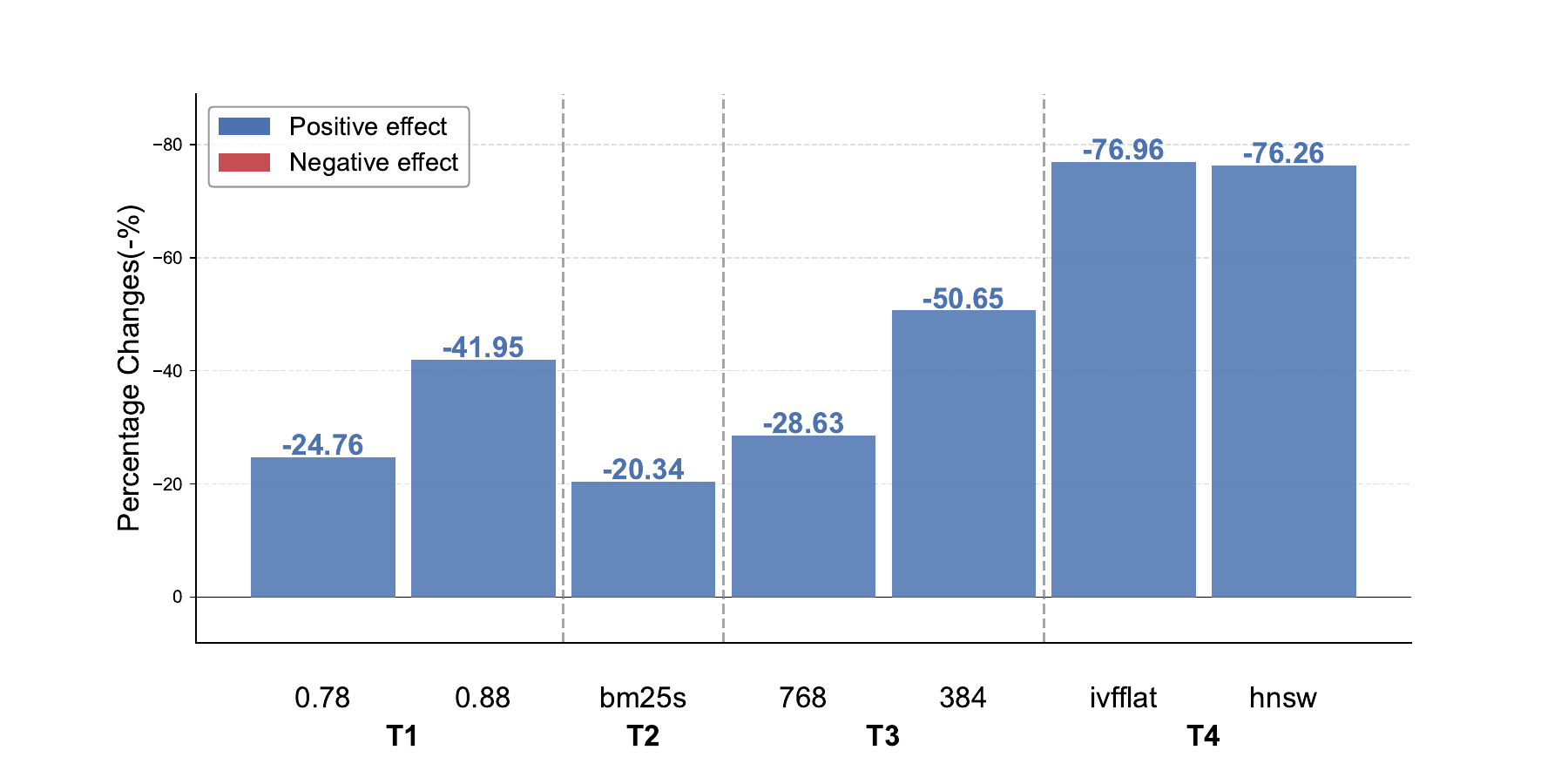}
        \caption{Relative Latency Difference to Baseline for Significant Techniques}
        \label{fig:percent_change_latency}
    \end{subfigure}

    \caption{Latency Experiment Results}
    \label{fig:latency_results}
\end{figure}

In this section, we examine the impact of the evaluated techniques on system latency and explore the relationship to the energy consumption results.
Fig.~\ref{fig:t_test_latency} shows the t-test results for latency.
We observe the same trend as for energy consumption:
all seven techniques that significantly reduced energy consumption also significantly reduced latency, with the exceptions again being T1 with the 0.68 threshold and T5.
However, the strengths of the reductions differed notably between the two QAs, highlighting that energy consumption and latency are not in a perfect linear relationship for complex distributed RAG systems.

For T1, as the threshold increased to 0.78 and 0.88, latency improved by 24.8\% and 42.0\%, respectively.
The 0.68 threshold was again too similar to the baseline to lead to significant changes.

Regarding T2 (the BM25S reranker), the improvement in latency was less pronounced compared to the improvement in energy consumption (20.3\% vs. 32.6\%).
This suggests that the additional response time introduced by the reranker component offset some of the overall latency gains, but its filtering capability still led to more substantial downstream energy savings.

For T3, embedding size reductions to 768 and 384 improved latency by 28.6\% and 50.7\% respectively.
Notably, going from the 1024 dimensions of the baseline to 384 led to a greater reduction in latency than increasing the T1 threshold from 0.58 to 0.88, despite the 0.88 threshold T1 variant yielding more substantial energy reduction.
This again highlights that using latency reductions as a proxy for energy reductions is not always reliable.

Moreover, the T4 indexing strategies again yielded the most substantial improvement in latency among all techniques.
Both IVFFlat and HNSW resulted in similar latency reductions of about 76\% compared to the baseline.

Lastly, T5 was similarly ineffective for latency reductions as it was for energy consumption.
The reason was likely the same: used queries and documents were not similar enough to make good use of the caching functionality.

\subsection{Trade-off With Accuracy (RQ2b)}
In this section, we first examine the baseline's ability to answer questions accurately and then evaluate the impact of the proposed techniques on system accuracy.
Among all 2,606 questions from our benchmark dataset, the baseline system produced, on average, correct answers for 619 questions, resulting in an accuracy of 23.75\%.
This is in line with the published CRAG benchmark results~\cite{crag_benchmark}, which mention an accuracy of 23.7\% for Llama 3.1 8B Instruct.
Our baseline system achieved almost identical accuracy, which validates the realism and effectiveness of our base setup, but also of our LLM-as-Judge approach.

\begin{figure}
    \centering
    \begin{subfigure}{\linewidth}
        \includegraphics[width=\linewidth]{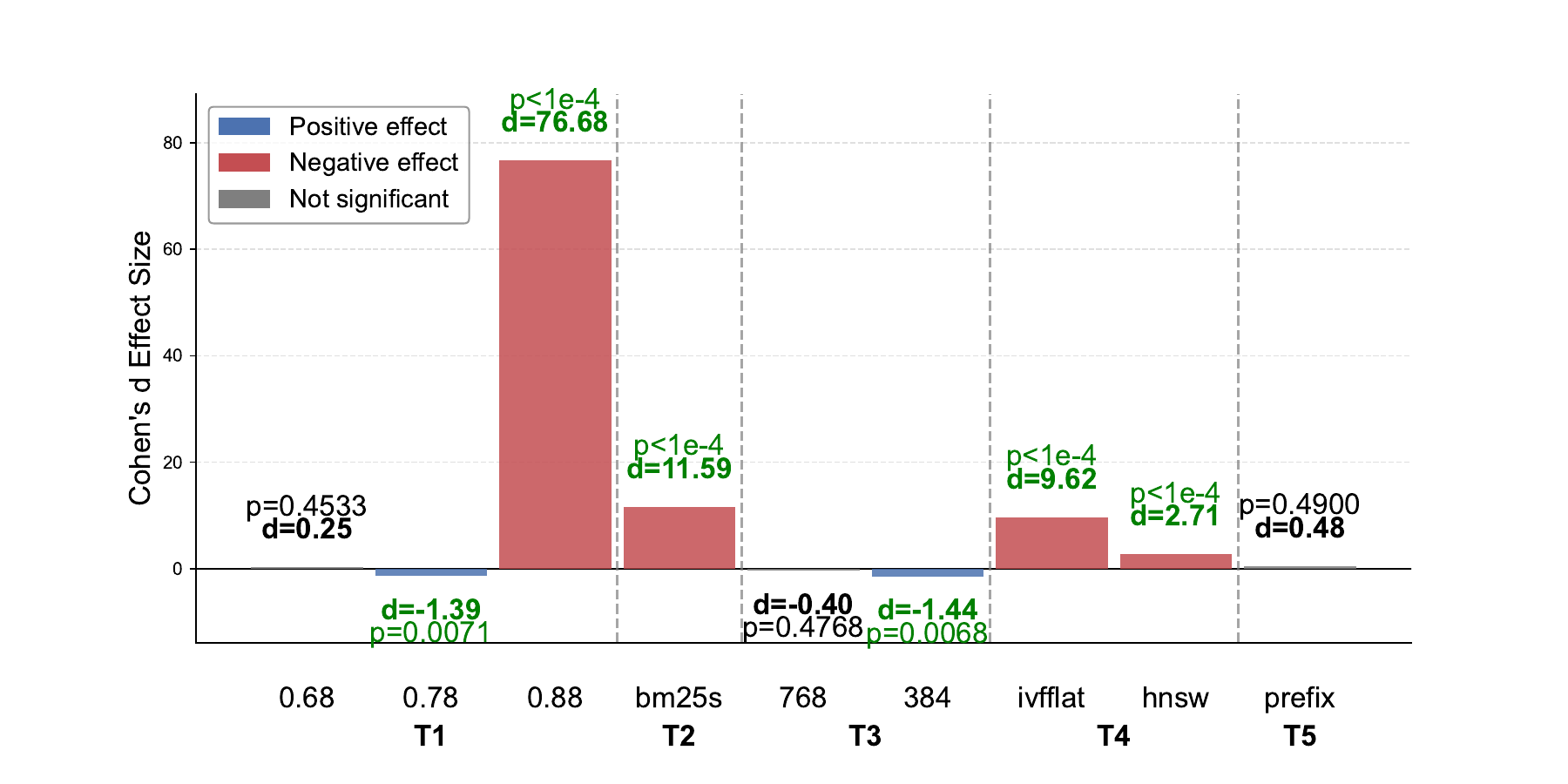}
        \caption{T-Test Results for Accuracy}
        \label{fig:t_test_accuracy}
    \end{subfigure}

    \vspace{0.0em}

    \begin{subfigure}{\linewidth}
        \includegraphics[width=\linewidth]{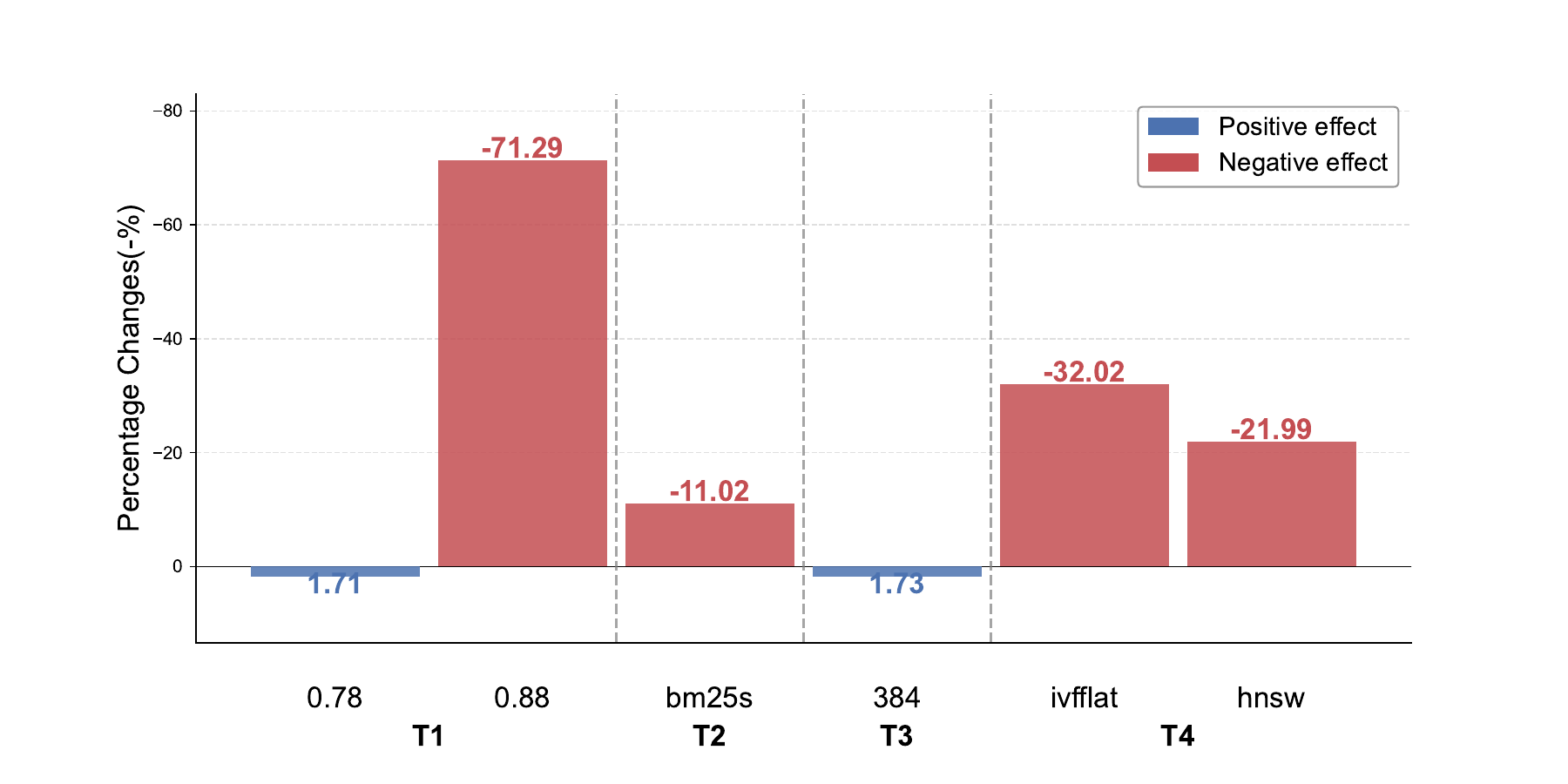}
        \caption{Relative Accuracy Difference to Baseline for Significant Techniques}
        \label{fig:percent_change_accuracy}
    \end{subfigure}

    \caption{Accuracy Experiment Results}
    \label{fig:accuracy_results}
\end{figure}

We show the t-test results for accuracy in Fig.~\ref{fig:t_test_accuracy} and the respective percentage change of significant techniques in Fig.~\ref{fig:percent_change_accuracy}.
For T1, increasing the retrieval threshold to 0.68 and 0.78 did not significantly harm accuracy, with the latter even significantly improving it by 1.7\%.
However, the 0.88 threshold had a substantial negative impact, with accuracy dropping by a massive 71.3\%.
When the threshold is low, the system retains more data points, which helps maintain accuracy but limits improvement on energy consumption and latency.
As the threshold increases, fewer data points are used, improving efficiency at the cost of potentially discarding important information, which explains the sharp drop in accuracy at 0.88.
This highlights a promising direction for future research:
efficiently identifying the \enquote{golden threshold} that optimally balances energy consumption and latency without compromising accuracy for a given RAG system.
Since T1 with the 0.78 threshold also contributed to improvements in energy consumption and latency, it was one of the few techniques with advantageous outcomes for all three QAs.

Regarding T2 with the BM25S reranker, its observed improvements in both latency and energy consumption unfortunately came at the cost of accuracy, with a decrease of approximately 11\%.
This suggests that the reranker may filter out some critical documents retrieved by pgvector that are essential for generating accurate responses.
As a result, the final set of documents passed to the language model is less informative, leading to a reduction in overall accuracy.
This trade-off might be acceptable for some practical RAG use cases, while it may disqualify the technique for many others.

For T3, reducing the embedding size to 768 did not significantly affect accuracy.
Additionally, when reducing the embedding size further to 384, there was even a slight but significant increase in accuracy (1.7\%).
As discussed earlier, both T3 variants also significantly improved energy consumption and latency, making T3 the second technique with no drawbacks for our three QAs.
This suggests that decreasing the embedding size did not compromise the model’s ability to retrieve relevant documents from pgvector.
One possible explanation is that the queries in our dataset are relatively short, resulting in the system consistently retrieving the full top-$k$ documents (150).
Consequently, the retrieved document set and its size remained stable across embedding sizes, which might have been different for longer queries.

While both T4 indexing strategies had substantial benefits for energy consumption and latency, they unfortunately also significantly reduced accuracy.
Both IVFFlat and HNSW had a very similar effect on the first two QAs, but their negative impact on accuracy notably differed.
The use of HNSW resulted in an accuracy decrease of 22.0\%, whereas IVFFlat led to an even larger drop of 32\%.
This suggests that, under default parameter settings, HNSW may be the better choice when balancing energy consumption and accuracy, but that both are still  unlikely to be usable in practice:
their accuracy drops will simply be unacceptable for most RAG use cases.

Lastly, the prefix caching of T5 did not significantly impact accuracy, similar to the other two QAs, making it the least impactful technique in our experiment.

\begin{table*}
    \centering
    \small
    \renewcommand{\arraystretch}{0.9}
    \caption{Combined Experiment Results for All Dependent Variables}
    \label{tab:hypothesis-results}
    \begin{tabular}{lllrrr}
        \textbf{Technique} & \textbf{Config} & \textbf{Metric} & \textbf{$\Delta$ (\%)} & \textbf{p-value} & \textbf{Cohen's d} \\
        \midrule
        \midrule
        \multirow{9}{*}{T1} & \multirow{3}{*}{0.68} & Energy Consumption & --     & 0.66219 & -- \\
                             &                        & Latency            & --     & 0.64539  & -- \\
                             &                        & Accuracy           & --     & 0.42672  & -- \\
                             \cmidrule(lr){2-6}
                             & \multirow{3}{*}{0.78} & Energy Consumption & -20.68 & 0.00107  & 1.657 \\
                             &                        & Latency            & -24.76 & 0.00207  & 1.664 \\
                             &                        & Accuracy           & 1.71   & 0.00062  & -2.026 \\
                             \cmidrule(lr){2-6}
                             & \multirow{3}{*}{0.88} & Energy Consumption & -51.02 & <0.0001  & 4.052 \\
                             &                        & Latency            & -41.95 & <0.0001  & 2.919 \\
                             &                        & Accuracy           & -71.29 & <0.0001  & 79.965 \\
        \midrule
        \multirow{3}{*}{T2} & \multirow{3}{*}{BM25S} & Energy Consumption & -32.64 & <0.0001 & 2.425 \\
                             &                        & Latency            & -20.34 & 0.00831 & 1.289 \\
                             &                        & Accuracy           & -11.02 & <0.0001 & 10.235 \\
        \midrule
        \multirow{6}{*}{T3} & \multirow{3}{*}{768}  & Energy Consumption & -20.00    & 0.00173 & 1.48 \\
                             &                        & Latency            & -28.63 & 0.00204 & 1.722 \\
                             &                        & Accuracy           & --     & 0.75527 & -- \\
                             \cmidrule(lr){2-6}
                             & \multirow{3}{*}{384}  & Energy Consumption & -38.12 & <0.0001 & 2.964 \\
                             &                        & Latency            & -50.65 & <0.0001 & 3.567 \\
                             &                        & Accuracy           & 1.73   & 0.01490 & -1.227 \\
        \midrule
        \multirow{6}{*}{T4} & \multirow{3}{*}{IVFFlat} & Energy Consumption & -59.76 & <0.0001 & 4.806 \\
                             &                          & Latency            & -76.96 & <0.0001 & 5.680 \\
                             &                          & Accuracy           & -32.02 & <0.0001 & 10.342 \\
                             \cmidrule(lr){2-6}
                             & \multirow{3}{*}{HNSW}   & Energy Consumption & -58.63 & <0.0001 & 6.061 \\
                             &                          & Latency            & -76.26 & <0.0001 & 6.594 \\
                             &                          & Accuracy           & -21.99 & <0.0001 & 2.677 \\
        \midrule
        \multirow{3}{*}{T5} & \multirow{3}{*}{Prefix caching} & Energy Consumption & --     & 0.95776 & -- \\
                             &                               & Latency            & --     & 0.09637 & -- \\
                             &                               & Accuracy           & --     & 0.26845 & -- \\
        \midrule
        \midrule
    \end{tabular}
\end{table*}

\begin{figure}[h]
    \includegraphics[width=\linewidth]{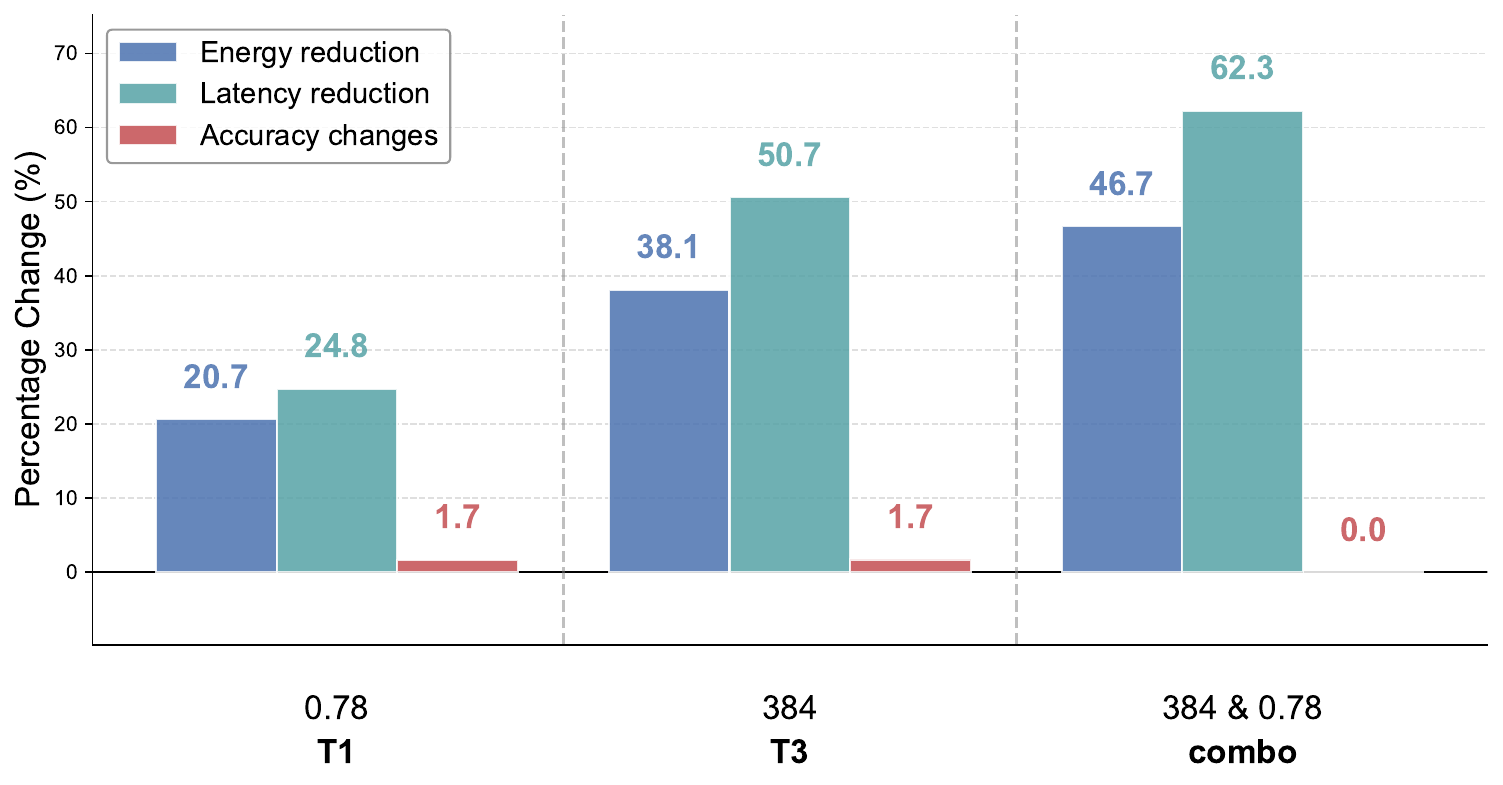}
    \caption{Relative Differences for T1-0.78 and T3-384 Combination Compared to Baseline}
    \label{fig:combo_percentage_changes}
\end{figure}

\subsection{Combining the Two Best Techniques}
\label{s:results_combination}
Two technique configurations (T1 with the 0.78 threshold and T3 with 384 dimensions) emerged as the most beneficial candidates, improving energy consumption, latency, and even slightly accuracy.
To understand whether their benefits compound when applied together, we ran an additional experiment with a system variation that implemented both of them simultaneously.
As shown in Fig.~\ref{fig:combo_percentage_changes}, the combination of the two led to even greater improvements in energy consumption and latency.
Specifically, T1-0.78 alone achieved a 20.7\% reduction in energy consumption, while T3-384 alone yielded a 38.1\% reduction.
Their combination, however, reduced energy consumption by 46.7\%.
The results for latency were similarly beneficial.
Although each individual configuration results in a modest accuracy improvement of 1.7\%, their combination did not produce a statistically significant change in accuracy anymore.
According to \citet{gu2021principled}, as the dimensionality of embeddings increases, the distribution of pairwise cosine similarities tends to converge toward a stable distribution with finite variance.
In higher-dimensional spaces, cosine similarity scores cluster more tightly around a central value, reducing the variance between similarity scores across different pairs.
In practice, high-dimensional embeddings tend to make random vectors more orthogonal, concentrating pairwise cosine similarities near zero with low variance.
This shifts the natural baseline similarity upward, e.g., from a slightly negative value in lower dimensions to near zero, which likely requires a threshold recalibration.
Nonetheless, since the combination also did not harm accuracy, applying both techniques at once remains a powerful option to improve energy efficiency.

\section{Discussion}
\label{s:discussion}
In this section, we summarize the key findings of our study, provide additional explanations, and discuss their implications.
Table~\ref{tab:hypothesis-results} provides an aggregated summary of the impact of each technique on the studied dependent variables, which serves as a quick reference for identifying the most effective techniques and understanding their associated trade-offs.

\begin{figure}
    \includegraphics[width=\linewidth]{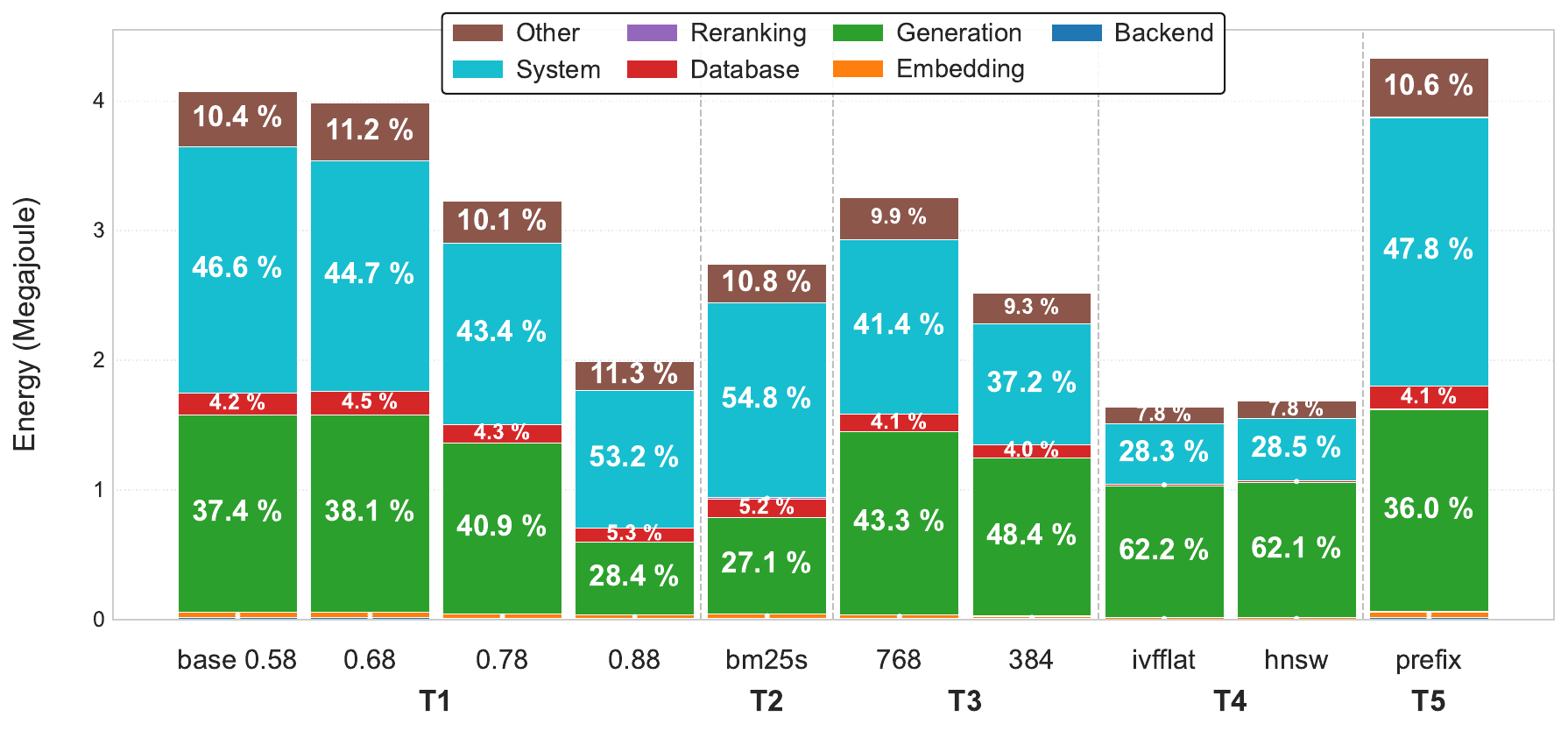}
    \caption{Energy Consumption Breakdown by Component}
    \label{fig:energy_breakdown}
\end{figure}

As shown in our results, \textbf{applying indexing strategies (T4) led to the strongest energy decreases}, with both IVFFlat and HNSW indexing achieving reductions of about 59\%.
One reason for that is that the energy consumption of the database dropped substantially due to the indexing (see Fig.~\ref{fig:energy_breakdown}).
In all other configurations, the database accounts for more than 4\% of total energy usage, whereas this share is noticeably reduced when indexing is applied.
Additionally, system-level processes like background services and OS tasks also consumed less energy.
This indicates that indexing enabled faster and more efficient similarity searches, lowering query latency and CPU usage for the database itself while also reducing overall system resource contention.
But despite their energy benefits, indexing comes with a trade-off: since they also reduce the number of retrieved documents, both IVFFlat and HNSW decrease accuracy by more than 20\%, which is likely unacceptable in practice.
However, it is important to note that we did not explore different parameter configurations for IVFFlat.
For example, increasing the \texttt{probes} parameter is known to improve recall, potentially mitigating some accuracy loss.
Future research could further investigate the trade-offs of different IVFFlat settings to better optimize energy consumption without sacrificing accuracy.

Another major result was that \textbf{similarity threshold increases (T1) and embedding size reductions (T3) were the most beneficial techniques}.
Unlike other techniques, these two significantly reduced energy consumption without sacrificing latency or accuracy.
In fact, each of them led to a modest accuracy improvement of 1.7\%.
This suggests that RAG systems should carefully determine an optimal threshold based on dataset sampling and select an appropriate embedding model for balanced system performance.
Interestingly, the combination of T1 and T3 returned the accuracy to the baseline.
This finding reveals that \textbf{the optimal similarity threshold in RAG systems is not fixed but can shift depending on embedding configurations.}
Therefore, when replacing a high-dimensional embedding model with a lower-dimensional one to reduce energy and latency, it is crucial to recalculate the similarity threshold to preserve potential accuracy gains.

Moreover, our experiment showed that \textbf{controlling the overall content size is critical for improving energy efficiency in enterprise RAG systems.}
As shown in Fig.~\ref{fig:energy_gpu_percentage}, both threshold adjustment and indexing strategies significantly reduced GPU energy consumption.
Since the LLM itself remains fixed on the GPU, these results demonstrate that such techniques can effectively regulate the amount of data flowing into the generation model, a stage that accounts for a substantial portion of energy consumption (see Fig.~\ref{fig:energy_breakdown}).
This highlights the importance of our study, especially in real-world scenarios where frozen LLMs are commonly used.

Finally, our results highlight that reducing the energy consumption in RAG systems is not difficult, as 7 of 9 treatments achieved this.
However, only two techniques could do so without unacceptable trade-offs.
This underscores that the real challenge in Green AI is \textbf{achieving energy efficiency}, i.e., reducing energy consumption without harming other important QAs, especially accuracy.
Future research will have to focus on such techniques, as practitioners will not adopt techniques with substantial drawbacks.

\section{Threats to Validity}
\label{sec:threats}
We assess potential threats to the validity of our experiment following the framework of Wohlin et al.~\cite{wohlin12}.

\textbf{Internal Validity.}
Chunking is critical for RAG systems, as larger chunks can improve recall but may reduce precision~\cite{pesl2025retrievalaugmentedgenerationservicediscovery}.
We used the \texttt{TokenTextSplitter} of Spring AI with a chunk size of 320 tokens to balance model compatibility and manageable chunk counts.
Despite this, 10 out of 501,916 chunks failed, which could marginally affect accuracy if they contained relevant information.

Experiments were conducted on a university lab machine with controlled access via a shared Google calendar and a monitoring script that automatically logged out unauthorized users.
Additionally, we stopped unnecessary background processes before the experiment.
Nevertheless, some residual background processes cannot be fully excluded.
Environmental factors, such as room temperature, may also affect components sensitive to thermal throttling, such as GPUs.
To reduce the impact of potential short-term fluctuations, we ran each treatment 10 times with sufficient duration to allow stabilization.
We also executed a warm-up dataset of 100 queries before each trial to further stabilize the system.
We therefore believe our energy and latency measurements to be fairly reliable.

LLM-as-Judge approaches enable efficient accuracy evaluations of complex NLP tasks, but they are also subject to potential LLM hallucinations.
While our DeepSeek-V2 judge reported a similar accuracy for the baseline system as previous benchmarks, it is possible that some answers were evaluated wrongly.
Overall, we still believe that the accuracy changes between treatments are reliable, even though the absolute numbers could be slightly different.

\textbf{External Validity.}
Our setup mirrored a production RAG system at SIG, and our close collaboration ensured realistic design choices.
However, alternative implementations exist, such as running all components in a single process, as in the CRAG benchmark.
Experiments were conducted in a controlled, on-premise environment with fixed hardware (24 GB VRAM GPUs), which may limit generalization to cloud or distributed deployments, where features like microservice autoscaling can improve latency.
While our results should be transferable to similar chatbot RAG systems, we have to be careful with broader generalization.

\textbf{Conclusion Validity.}
To control Type I errors from multiple comparisons, we applied the Holm-Bonferroni correction.
Each treatment was also executed 10 times, yielding stable, normally distributed measurements.
While more trials could improve stability, the current setup was sufficient to support valid conclusions.

\section{Conclusion}\label{s:conclusion}
In this paper, we provided a thorough analysis of proposed techniques that could reduce the energy consumption of RAG systems while also studying potential trade-offs with latency and accuracy.
We identified valuable findings for researchers and practitioners, such as the importance of selecting an optimal threshold in the retrieval stage.
An appropriately chosen similarity threshold (T1) can significantly reduce energy consumption and latency without compromising accuracy.
However, setting the threshold too high may severely degrade answer quality.
Similarly, using smaller embedding models (T3) is beneficial when handling short queries and a large number of retrieved documents, offering energy and latency benefits with no loss in accuracy and, in some cases, even a slight increase.
We also showed that applying indexing strategies like IVFFlat and HNSW in pgvector (T4) or integrating a lightweight reranker (T2) can significantly reduce energy consumption and latency.
Nonetheless, these enhancements come at the cost of reduced accuracy, highlighting a crucial trade-off that future research must continue to explore.
Our results can support RAG practitioners in substantially optimizing energy consumption, which has important societal implications.

Since the energy efficiency of RAG systems is a relatively new research area, future work should build upon our findings to explore this important topic further.
For instance, examining combinations of existing techniques may reveal new patterns of interaction.
This could lead to the discovery of more effective configurations that optimize energy usage without sacrificing system performance or accuracy.
To support such studies, we make all experimental artifacts and implementation details publicly available.\footnote{\url{https://doi.org/10.5281/zenodo.17990845}}

\begin{acks}
This experiment was conducted with the support of the Green Lab at VU Amsterdam and additional colleagues at SIG. We gratefully acknowledge everyone’s assistance with this study.
\end{acks}

\bibliographystyle{ACM-Reference-Format}
\bibliography{references}

\end{document}